\documentstyle[prb,aps,epsfig]{revtex}

\begin{document}

\title{Charge spectrometry with a strongly coupled
superconducting single-electron transistor}
\author{C. P. Heij, P. Hadley, and J. E. Mooij}
\address{Applied Physics and DIMES, Delft University of Technology\\
Lorentzweg 1, 2628 CJ Delft, The Netherlands}
\date{\today}
\maketitle

\begin{abstract}
We have used a superconducting single-electron transistor as a
DC-electrometer that is strongly coupled to the metal island of
another transistor. With this set-up, it is possible to directly
measure the charge distribution on this island. The strong
capacitive coupling was achieved by a multilayer fabrication
technique that allowed us to make the coupling capacitance bigger
than the junction capacitances. Simulations of this system were
done using orthodox theory of single-electron tunnelling and
showed excellent agreement with the measurements.
\end{abstract}

\pacs{73.23.Hk, 85.30.Wx, 85.25.Na}

A single-electron transistor can make extremely sensitive charge
measurements with a resolution of about
$10^{-5}$~$e/\sqrt{\text{Hz}}$.
\cite{krupenin00,schoel98,visscher96} In principle, this
sensitivity is sufficient to measure the charging and discharging
of a small conductor as current flows through that conductor. A
single-electron transistor would be able to register the
tunnelling of individual electrons tunnel onto the conductor from
one lead and off the conductor to another lead for currents up to
a few picoamp\`eres. In practice, the high output impedance of a
single-electron transistor makes this measurement very difficult.
If the charge on the conductor as a function of time could be
measured, then it would be possible to determine the occupation
probabilities of the various charge states. In charge state
$\langle 0 \rangle$ the conductor is neutrally charged, in charge
state $\langle 1 \rangle$ the conductor has an excess charge of
one electron, and so forth. The charge state occupation
probabilities play a central role in the theory of single-electron
tunnelling \cite{averin91a,ingold92} but are typically not
directly accessible experimentally. Here we report on an
experiment where a superconducting single-electron transistor
(SSET) was used to directly measure the charge state occupation
probabilities on the island of another nearby SSET.

An essential feature of this experiment was that the measurement
SSET (the electrometer) was strongly coupled to the island where
the current was flowing through. Strong coupling means that the
coupling capacitance was comparable to the total capacitance of
the nearby island. When one electron was added to this island,
about $e/4$ of charge was induced on the coupling capacitor. The
electrometer was biased at a small voltage and it's tuning gate
was used to scan the charge on the neighboring island. These gate
traces directly reflect the charge distribution on this island.
The measurements are consistent with orthodox theory and they show
that a strongly coupled SSET can be used to directly measure the
charge distribution.

\section{Experiment}
\begin{figure}[tbh]
\centering\epsfig{figure=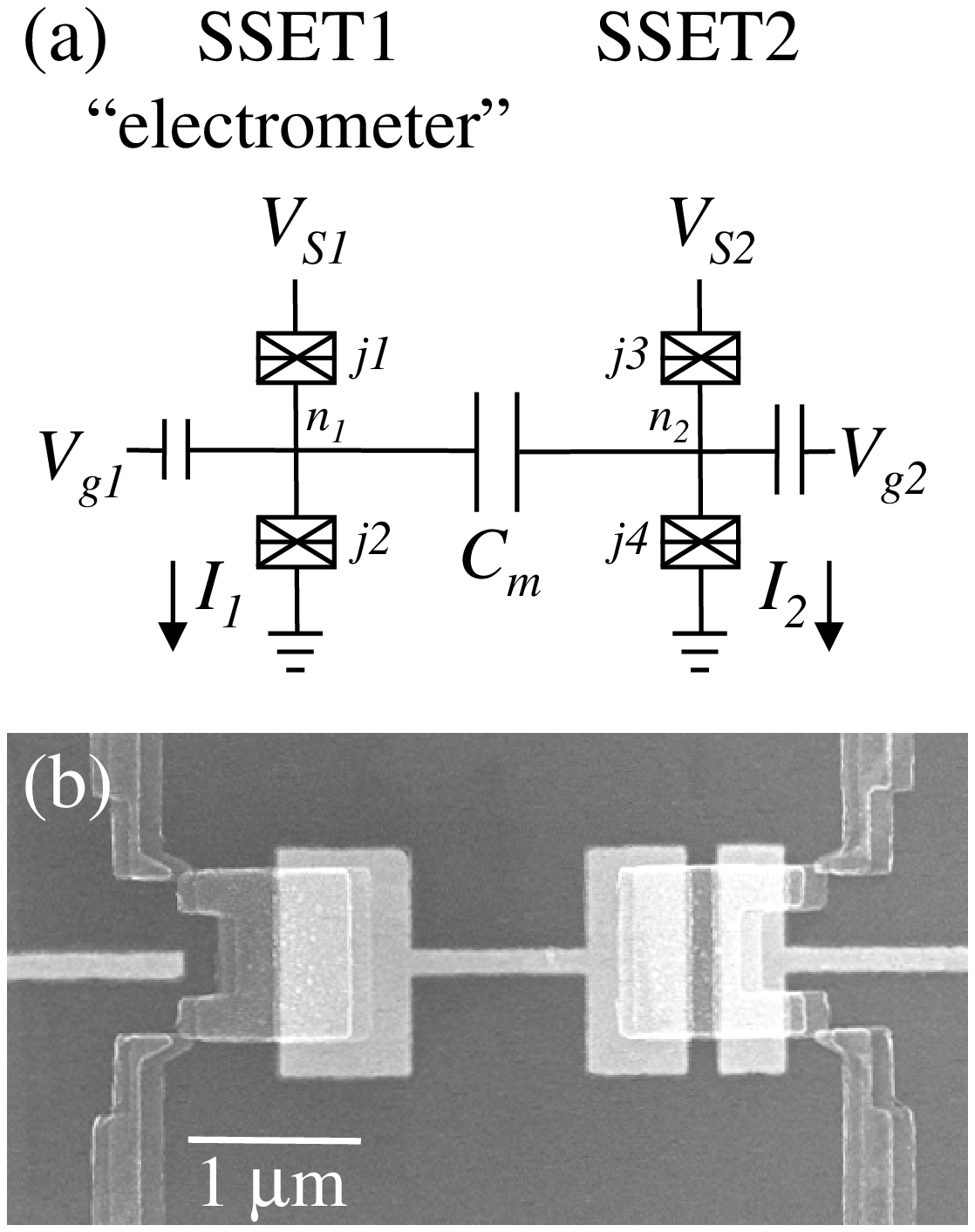, width=10cm, clip=true}
\caption{(a) Schematic of the circuit. The two junctions on the
left form the electrometer (SSET1), whose island is coupled
capacitively to the island of a nearby SSET (SSET2). (b) Scanning
electron microscope picture of the completed device. The light
gray layer is fabricated in gold, the aluminum layer shows up as
dark grey.} \label{circuit2sets}
\end{figure}

A schematic of the device is shown in Fig.~\ref{circuit2sets}a.
The device was fabricated in three layers. The junctions were
fabricated using standard shadow evaporation of aluminum. SSET1
has a planar gate capacitor $C_{g1}$ while the gate capacitor
$C_{g2}$ is defined as a parallel plate capacitor. Details of the
fabrication of a similar device were described elsewhere.
\cite{heij99} Figure \ref{circuit2sets}b shows a SEM picture of
the device. The two square islands of the SSETs are coupled via an
underlying dumbbell shaped coupling capacitor. The total effective
capacitance between the two islands is called $C_{m}$. Both SSETs
were biased asymmetrically, connected to a voltage source at one
side and grounded at the other side. The device was measured in a
dilution refrigerator with a base temperature of 10 mK. The leads
were equipped with $\pi$-filters at room temperature and standard
copper-powder filters at base temperature. The effective electron
temperature, 25 mK, was measured in the normal state by fitting
experimentally obtained Coulomb peaks. All further measurements
were done in the superconducting state, the superconducting gap
being $\Delta=200~\mu e$V.

\begin{figure}[tbh]
\centering\epsfig{figure=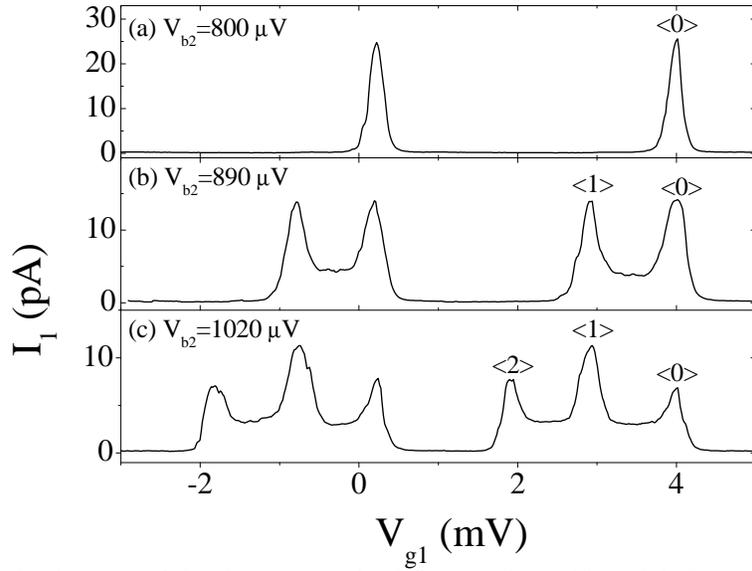, width=10cm, clip=true}
\caption{Experimental Coulomb traces of the electrometer for
different values of $V_{b2}$ while $V_{b1}=5 \mu$V,
$V_{g2}=10~\mu$V and T=25 mK. The extra peaks Coulomb peaks in (b)
and (c) correspond to the presence of extra electrons on island
2.} \label{sivgs}
\end{figure}

Throughout the measurements, the voltage bias of the electrometer
was kept constant at $V_{b1}= 805~\mu $V, just above $4\Delta/e$.
The current through the electrometer $I_1$ was measured as a
function of the gate voltage $V_{g1}$ and the bias voltage
$V_{b2}$ of the SSET2. The gate voltage $V_{g2}$ was kept
constant. Figure \ref{sivgs}a shows typical Coulomb oscillations
of the current through the electrometer. The gate of the
electrometer was swept while SSET2 was biased at 800 $\mu $V, the
current $I_2$ being negligibly small. Figure~\ref{sivgs}b shows
the same Coulomb trace when $V_{b2}=890~\mu $V, above the
quasiparticle threshold of SSET2. Surprisingly, the Coulomb peak
is split into two peaks, while at even higher bias
($V_{b2}=1090~\mu$V) it is split into three. As we will explain
below, each extra peak can be attributed to the presence of an
extra electron on the second island.

When $V_{b2}=800~\mu$V, the current through SSET2 is still
negligibly small and the occupation probability of charge state
$\langle 0 \rangle$ on island 2 is nearly 1. When $V_{b2}$ is
higher than the threshold voltage, a quasiparticle current $I_{2}$
will start to flow and the charge $n_2$ on the island 2 will
switch between $\langle 0 \rangle$ and $\langle 1 \rangle$. The
presence of an extra electron on island 2 will induce a fraction
of an electron on coupling capacitor $C_{m}$. By writing down the
total charge on both islands as a function of the capacitances and
the island potentials, one can show that this fraction is
$C_{m}/C_{\Sigma2}$, where $C_{\Sigma2}$ is the sum of all
capacitors connected directly to island 2. Table~\ref{table1}
gives the capacitance and resistance values of all the circuit
elements. With this table, we can calculate that the charge
induced on capacitor $C_{m}$ is 0.27 $e$. This results in the
extra Coulomb peak (labelled $\langle 1 \rangle$) shifted
$-e/C_{g1}*0.27=-1.04$ mV with respect to the peak labelled
$\langle 0 \rangle$ in Fig.~\ref{sivgs}b. When $V_{b2}$ is
increased even more, charge state $\langle 2 \rangle$ is also
populated on island 2 and three peaks appear (Fig.~\ref{sivgs}c).
Figure \ref{stabm} shows the Coulomb traces of the electrometer
current $I_{1}$ in grayscale versus the bias $V_{b2}$. One can
clearly see that consecutive charge states become populated with
increasing $V_{b2}$. The average bias voltage difference between
successive charge states on island 2 is $2E_{C2}/e$, where
$E_{C2}$ is the charging energy of island 2. The charge state
$\langle 4 \rangle$ becomes populated at $V_{b2}=1180~\mu$V and
induces 1.08 $e$ on island 1. The corresponding peak in the
Coulomb trace of the electrometer overlaps with the next set of
Coulomb peaks, limiting the number of observable charge states to
four with these circuit parameters.

By measuring Coulomb oscillations, the electrometer can be used to
directly resolve the average population of charge states on a
nearby island, even though the charge on this island changes on a
nanosecond timescale. At Coulomb peak $\langle j \rangle$, current
can only flow through the electrometer when the island 2 is in
charge state $\langle j \rangle$. The fraction of the time that
island 2 spends in charge state $\langle j \rangle$ is equal to
the relative peak height defined by,
\begin{equation}
p_{i}=\frac{I_{p,i}}{\sum_j I_{p,j}},  \label{occup}
\end{equation}
where $I_{p,j}$ is the peak height of the Coulomb peak
corresponding to the charge state $\langle j \rangle$ on island 2.
Simulations confirm that the height of the individual peaks
reflects the exact population of the corresponding charge state.

The relative peak heights of the Coulomb traces in
Fig.~\ref{sivgs}b,c are calculated and shown in column I of
Table~\ref{table2}. They closely match the occupation of the
various charge states as calculated in the simulations for
identical bias conditions, shown in column II. Also, for bias
conditions other than $V_{b2}=890~\mu$V and $1020~\mu$V, the
simulated relative peak heights closely match the experimental
ones. This shows that a strongly coupled SSET can be used to
quantitatively measure the charge {\it distribution} on a nearby
object.

\section{Simulations}
The current through both SSETs was calculated using a master
equation analysis. By solving the master equation we can calculate
the occupation probabilities of the various charge states. The
master equation for this two island system is,
\begin{equation}
\frac{\partial P_{ij}}{\partial t} =\sum_{kl\neq ij}\left(
P_{kl}\Gamma _{kl\rightarrow ij}-P_{ij}\Gamma _{ij\rightarrow
kl}\right), \label{master2a}
\end{equation}
\begin{equation}
\sum_{ij}P_{ij}=1, \label{master2b}
\end{equation}
where $P_{ij}$ is the probability that the system has $i$ excess
electrons on island 1 and $j$ excess electrons on island 2.
$\Gamma $ denotes the transition rate between different charge
states. In the stationary state, $P_{ij}$ does not change and the
left hand side of equation \ref{master2a} is zero. The first term
on the right describes the population of charge state $ij$ from
charge state $kl$ while the second term describes the depopulation
of charge state $ij$ to charge state $kl$. We neglect
co-tunnelling processes and $\Gamma $ only is non-zero when either
$i=k\pm 1$ or $j=l\pm 1$. Furthermore we only take into account a
maximum of five charge states per island.

The superconducting tunnel rates $\Gamma$ were then determined
with Fermi's Golden Rule using the superconducting density of
states and the free energy difference $\Delta F$ of a tunnelling
event. \cite{goldenrule,superdensity} $\Delta F$ is the sum of the
change in electrostatic energy plus the work done by the voltage
sources. The total electrostatic energy of the system can be
written as:
\begin{equation}
E(n_{1},n_{2})=E_{C1}( n_{01} + n_{1}) ^{2}+E_{C2}(n_{02}+n_{2})
^{2} + E_{m}( n_{01}+n_{1})(n_{02}+n_{2}), \label{ectotal}
\end{equation}
\begin{equation}
E_{C1} =\frac{e^{2}C_{2\Sigma }}{2(C_{\Sigma 1}C_{\Sigma
2}-C_{m}^{2})}=95~\mu e \text{V},  \label{ec1}
\end{equation}
\begin{equation}
E_{C2} =\frac{e^{2}C_{1\Sigma }}{2(C_{\Sigma 1 }C_{\Sigma 2
}-C_{m}^{2})}=53~\mu e \text{V},  \label{ec2} \end{equation}
\begin{equation}
E_{m} =\frac{e^{2}C_{m}}{C_{\Sigma 1}C_{\Sigma
2}-C_{m}^{2}}=51~\mu e \text{V},  \label{em}
\end{equation}
where $n_{01}$ and $n_{02}$ are the normalized charges induced on
the islands by the voltage sources, $n_{1}$ and $n_{2}$ are the
excess number of electrons on the islands and C$_{\Sigma 1}$ and
C$_{\Sigma 2}$ are the sum of all capacitors directly connected to
the respective islands. The electrostatic energy has three
contributions, the charging energies of the separate SSETs
(equations \ref{ec1} and \ref{ec2}) and the coupling energy
$E_{m}$, which describes the electrostatic interaction between
both SSETs. When equations \ref{master2a} and \ref{master2b} are
solved for $P_{ij}$, the total current $I_1$ can be calculated
with $P_{ij}$ and the tunnel rates.

\begin{figure}[tbh]
\centering\epsfig{figure=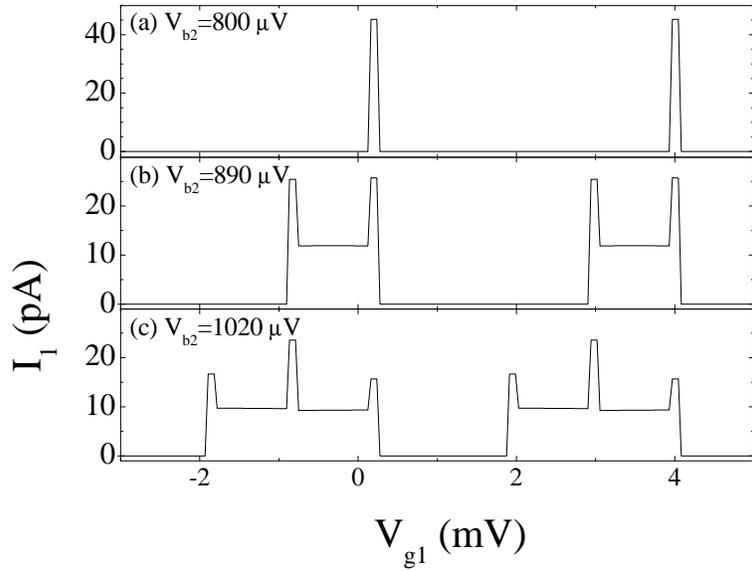, width=10cm, clip=true}
\caption{Simulations of Coulomb traces of the electrometer for
different values of~$V_{b2}$. $V_{b1}=805~\mu $V, $V_{g2}=10~\mu$V
and T=25 mK.} \label{ssivgs}
\end{figure}

In Fig.~\ref{ssivgs} the current through the electrometer has been
calculated for the same bias conditions as Fig.~\ref{sivgs}. One
can clearly see the extra Coulomb peaks appear when the bias
voltage $V_{b2}$ is increased. The absolute peak height of the
experiments is about 60\% of the peak height in the simulations.
This can be accounted for by the rounding of the superconducting
gap. Instead of the discontinuous jump in quasiparticle current
through a superconducting junction at $2\Delta/e$, in real
experiments, the current increases with a non-zero slope. In these
experiments, the differential resistance in this regime is about
5\% of the high bias junction resistance. For a bias voltage of
$V_{b1}=805~\mu $V, only $5~\mu $V above $4\Delta/e$, this has two
consequences. First, the Coulomb peaks have a more triangular form
as can be seen in Fig.~\ref{sivgs}, second, the Coulomb peak
height is smaller than in the simulations where the rounding has
not been taken into account. Simulations where the rounding of the
gap was taken into account with a simple model showed that the
rounding of the gap does not change the relative height of the
peaks, it merely decreases the overall current.

\begin{figure}[tbh]
\centering\epsfig{figure=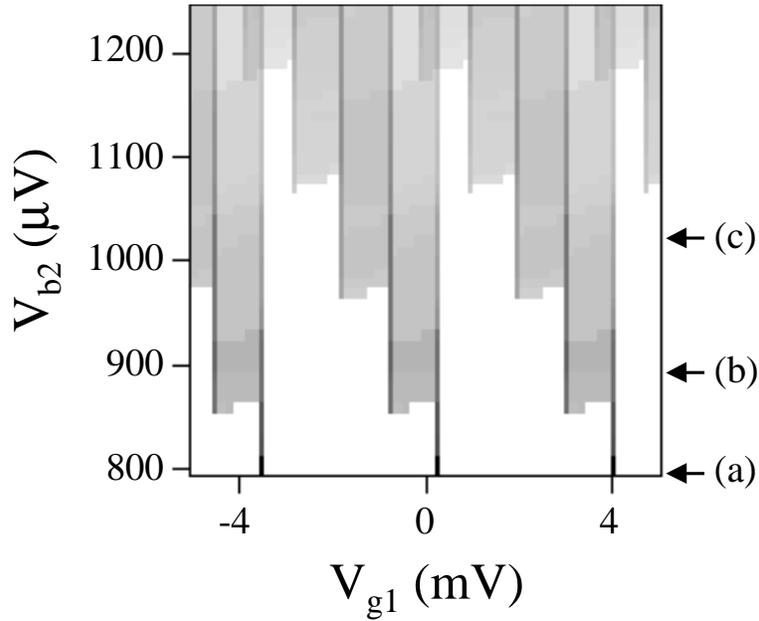, width=10cm, clip=true}
\caption{Simulation of the electrometer current versus $V_{g1}$
and $V_{b2}$. White indicates no current, black indicates a
current of 45 pA. The arrows indicate the values of $V_{b2}$ where
the traces of Fig.~\ref{ssivgs}a-c have been extracted. }
\label{stabs}
\end{figure}

For the simulated Coulomb traces of Fig.~\ref{ssivgs}, the
relative peak heights as specified by Equation \ref{occup} are
given in Table~\ref{table2}. The experiments closely match the
simulated values. The relative peak heights in simulations are
slightly different though from the occupation of the charge states
on island 2 when the electrometer is switched "off"
($V_{b2}=800~\mu$V). Column three of Table~\ref{table2} shows the
undisturbed occupancies of the three charge states as determined
from the population matrix $P_{ij}$. As can be seen from
Table~\ref{table2}, the bias of the electrometer has a small back
action on the occupation of charge states on island 2. For the
bias range of Fig.~\ref{stabs}, it can be shown that the back
action of this electrometer changes the occupancies of the various
charge states by a maximum of $5\%$.

\begin{figure}[tbh]
\centering\epsfig{figure=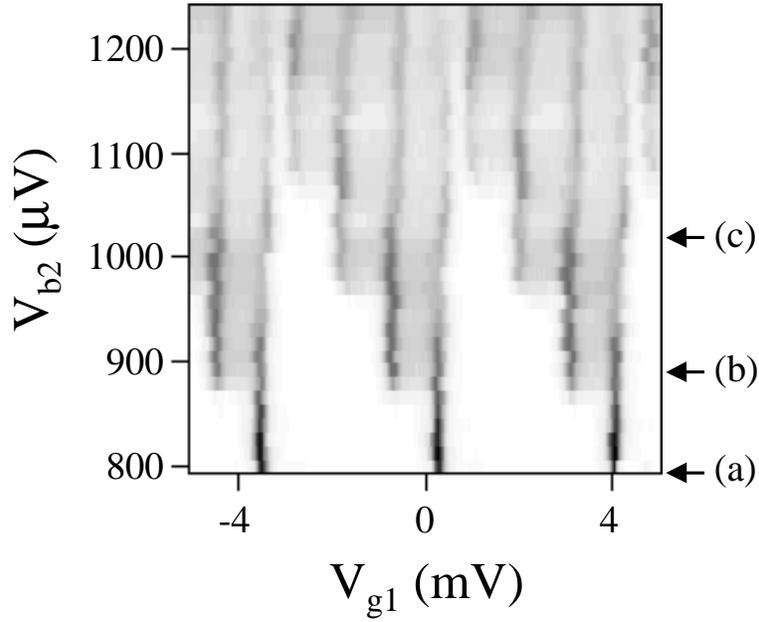, width=10cm, clip=true}
\caption{Electrometer current versus $V_{g1}$ and $V_{b2}$. White
indicates no current, black indicates a maximum current of 25 pA.
The arrows indicate the values of $V_{b2}$ where the traces of
Fig.~\ref{sivgs}a-c have been extracted. At $V_{b2}=1180~\mu$V the
charge state $\langle 4 \rangle$ becomes populated, but the
corresponding peak overlaps with the neighboring set of Coulomb
peaks.} \label{stabm}
\end{figure}

Figure \ref{stabs} is the simulated equivalent of
Fig.~\ref{stabm}. The extra peaks appear in Fig.~\ref{stabs} at
exactly the same bias conditions as in Fig.~\ref{stabm},
demonstrating the close agreement between experiments and
simulations. Another feature that clearly shows up in the
simulations as well as the measurements is the existence of a
current plateau in between the neighboring Coulomb peaks. Under
these bias conditions, the electron-tunnelling through both SSETs
is correlated. This effect has been described before in coupled 1D
arrays of tunnel junctions. \cite{averin91b,matters97} The details
of this effect will be discussed below.

\section{Discussion}
The ability to determine the position and the height of the extra
Coulomb peaks gives constraints on the bias conditions. In
general, the width of the peaks has to be smaller than the
separation between adjacent peaks. In the superconducting state
the width of Coulomb peaks is almost independent of temperature
for $k_{B}T<0.5 \Delta$ and depends linearly on the applied bias.
This constraint can be rewritten as:
\begin{equation}
eV_{b1}-4\Delta<E_{m}. \label{constraint}
\end{equation}
This simply states that the energy associated with the voltage
bias has to be smaller than the coupling energy. Because of the
quasiparticle threshold at $4\Delta/e$, this constraints the bias
voltage to $800~\mu\text{V} < V_{b1} < 851~\mu$V for this sample.
The quasiparticle rate is almost independent of the bias in this
bias window and simulations indicate that the back action of the
electrometer is also constant. If we take into account the
rounding of the gap and the experimental current noise,
$V_{b2}=805~\mu $V is about the optimal bias voltage, combining an
acceptable signal to noise ratio with a reasonably small width of
the Coulomb peaks. With the current sample parameters, we are
limited to the observation of a maximum of four charge states on
the neighboring island. We estimate that it is feasible to observe
at least seven different charge states, when the coupling
capacitance is lowered to $190$ aF, while keeping the other sample
parameters constant.

\begin{figure}[tbh]
\centering\epsfig{figure=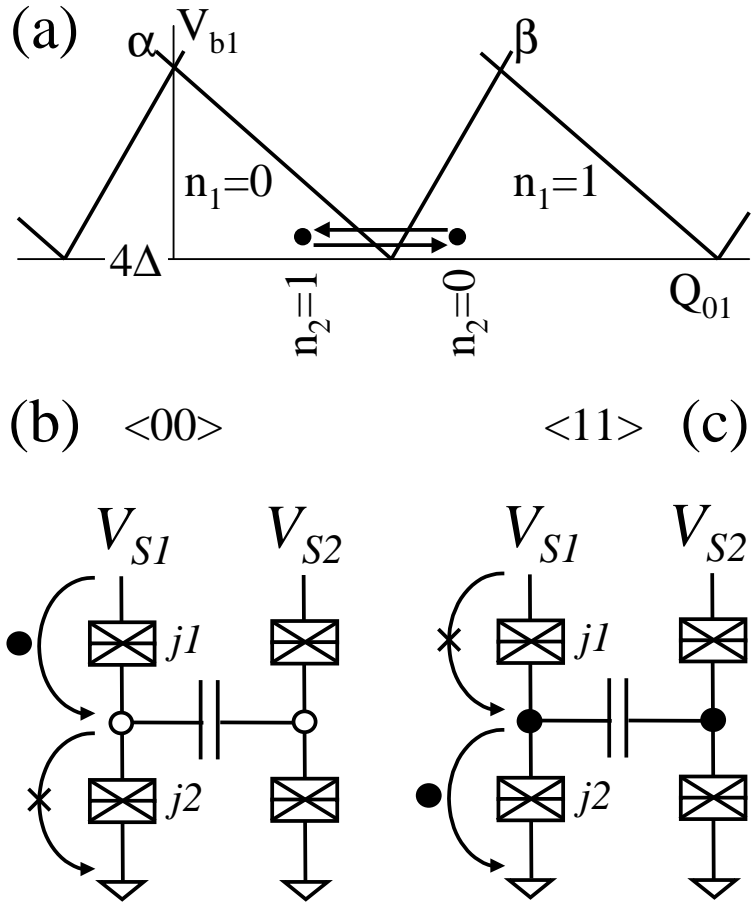, width=10cm, clip=true}
\caption{(a) Schematic of the quasiparticle thresholds of SSET1
above $V_{b1}=4\Delta/e$, shown as thick lines. The positions of
the dots denote the effective background charge induced by the
absence or presence of an extra electron on island 2. (b)
Visualization of the possible tunnelling events on island 1. When
the system is in charge state $\langle 00 \rangle$ it will decay
to $\langle 10 \rangle$ by an electron tunnelling through junction
j1. Electron tunnelling through junction j2 is energetically
unfavorable, just like electrons tunnelling upward. (c) In a
similar way charge state $\langle 11 \rangle$ decays to $\langle
01 \rangle$. } \label{diagram}
\end{figure}

Both Fig.~\ref{sivgs} and \ref{ssivgs} clearly show the existence
of a current plateau in between the accompanying Coulomb peaks. In
order to be able to measure the relative peak heights, this
plateau current should not exceed the Coulomb peak current and
therefore its mechanism should be understood. The mechanism can be
most easily explained when the number of occupied electron states
on island 2 is limited to two and under the assumption that the
tunnel rates in SSET1 are much larger than those in SSET2. Figure
\ref{diagram}a schematically displays the quasiparticle thresholds
for SSET1. The position of the dots denotes the effective
background charge when the charge state of island 2 is $\langle 0
\rangle$ (right dot) and $\langle 1 \rangle$ (left dot). The
position of the dots relative to each other is fixed. The bias
voltage $V_{b1}$ and hence the dots lie just above $4\Delta/e$.
With the gate voltage $V_{g1}$ the position of both dots can be
shifted along the $Q_{01}$ axis. If the gate voltage positions one
of the two dots above both quasiparticle thresholds $\alpha$ and
$\beta$ this leads to current in the form of a Coulomb peak. If
the dots are positioned as depicted in Fig.~\ref{diagram}a there
is an additional mechanism that will carry current.

The charge states with the lowest energy are now $\langle 10
\rangle$ and $\langle 01 \rangle$. If a current is forced to flow
through SSET 2 by biasing it above its quasiparticle threshold,
the following current cycle is most probable: If we start with
charge state $\langle 00 \rangle$ it is only favorable for
electrons to tunnel onto islands 1 or 2 via the top junctions.
Because we assume that the tunnel rates in SSET1 are much larger
than those in SSET2, an electron will most probably tunnel through
junction j1 first, as shown in Fig.~\ref{diagram}b. Now the system
is in the charge state $\langle 10 \rangle$ which is stable for
electron tunnelling in SSET1. After some time the bias voltage
$V_{b2}$ forces an electron on island 2 and the system is in state
$\langle 11 \rangle$. As can be seen in Fig.~\ref{diagram}c this
state decays to $\langle 01 \rangle$ through junction j2, again
assuming the tunnel rates are much higher in SSET1. This charge
state is also stable for electron tunnelling in SSET1. The cycle
is completed when the electron is forced off island 2 and the
system is back in charge state $\langle 00 \rangle$. The cycle of
one electron tunnelling through SSET2 has transported another
electron through SSET1 making $I_1=I_2$. This cycle is possible
for all gate voltages where both the position of $n_2=\langle 1
\rangle$ lies below quasiparticle threshold $\alpha$ and position
of $n_2=\langle 0 \rangle$ lies below quasiparticle threshold
$\beta$. This gives rise to a current plateau exactly in between
the Coulomb peaks attributed to the both charge states.

In this sample, the resistances of SSET1 and SSET2 and hence the
tunnel rates differ by only a factor of two. This means that
cycles can be missed, for example if state $\langle 00 \rangle$
decays to $\langle 01 \rangle$, the system is forced to $\langle
00 \rangle$. An electron is transported through SSET2, without
giving rise to current in SSET1. The general equation for the
relation between $I_1$ and $I_2$ can be deduced by analytically
solving the master equation under the assumption that only the
four charge states $\langle 00 \rangle$, $\langle 10 \rangle$,
$\langle 01 \rangle$ and $\langle 11 \rangle$ need to be
considered. If we assume that the tunnel rates through junction j1
and j2 are equal and called $\Gamma_1$, as well as those through
j3 and j4 are equal and called $\Gamma_2$ this yields:
\begin{equation}
I_{1}=\frac{\Gamma_{1}}{\Gamma_{1}+2\Gamma _{2}}I_{2}.
\label{copier}
\end{equation}
By deriving the expressions for the Coulomb peak current, it can
be shown that the peak currents are always sufficiently larger
than the plateau current, making it possible to adequately
determine the relative peak heights. When the number of occupied
charge states on island 2 is larger than two, the mechanism
leading to the current plateaus is similar, but different
combinations of charge states might be stable and equation
\ref{copier} will be modified. Again though, the plateau current
is always smaller than the Coulomb peaks adjacent to the
particular current plateau.

We also studied the performance of the electrometer in the normal
state. In the normal state, however, the Coulomb peaks are very
sensitive to thermal fluctuations. The thermal broadening of the
Coulomb peaks at 70 mK was enough to merge the adjacent Coulomb
peaks, making an accurate determination of the relative peak
heights impossible. Additionally, due to a mechanism similar to
the one leading to plateau current in the superconducting state,
the adjacent Coulomb peaks merged at 30 mK in the normal state,
making the normal state operation of this electrometer
impractical.

\section{Conclusions}
We have used a SSET to measure the charge distribution on a
neighboring island. Both the islands were strongly coupled by a
multi-layer technology. The presence of an extra electron on a
neighboring island split the Coulomb peaks of the SSET. The
relative height of these peaks directly translates to the
occupation of the associated charge state. In between the
neighboring Coulomb peaks the current is carried by correlated
tunnelling of electrons through both SSETs.

\section{Acknowledgements}
We thank K. K. Likharev for illuminating the basic mechanism
leading to the current plateau.

\newpage

\vspace{2cm}

\vspace{2cm}

\vspace{2cm}

\vspace{2cm}

\vspace{2cm}

\vspace{2cm}

\newpage

\begin{table}[tbh] \centering
\begin{tabular}{|l|l|l|l|l|l|l|l|l|l|}
\hline & j1 & j2 & j3 & j4 & C$_{g1}$ & C$_{g2}$ & C$_{m}$ & C$_{\Sigma 1}$ & C$_{\Sigma 2}$ \\
\hline C (aF) & 135 & 350 & 160 & 400 & 42 & 640 & 450 & 977 & 1650 \\
\hline R (M$\Omega )$ & 3.5 & 3.5 & 6.5 & 6.5 & $\infty$ & $\infty$ & $\infty$ & - & - \\
\hline
\end{tabular} \caption{Capacitance and resistance values for
the circuit parameters as calculated from the stability diagrams
and current-voltage characteristics of both SSETs.} \label{table1}
\end{table}

\begin{table}[tbh] \centering
\begin{tabular}{|l||l|l|l||l|l|l||l|l|l||}
\hline & \multicolumn{3}{c||}{I. experiments} &
\multicolumn{3}{c||}{II. simulations} & \multicolumn{3}{c||}{III. SSET1 'off'} \\
\hline $V_{b2}$ ($\mu $V) & $p_{0}$ & $p_{1}$ & $p_{2}$ & $p_{0}$
& $p_{1}$ & $p_{2} $ & $p_{0}$ & $p_{1}$ & $p_{2}$ \\ \hline 890 &
0.50 & 0.50 & 0 & 0.50 & 0.50 & 0 & 0.50 & 0.50 & 0 \\ \hline 1020
& 0.28 & 0.43 & 0.29 & 0.28 & 0.42 & 0.30 & 0.29 & 0.40 & 0.31 \\
\hline
\end{tabular} \caption{The population of first three charge states
on island 2, as calculated from the peak heights in the
experiments (Fig.~\ref{sivgs}) and the simulations
(Fig.~\ref{ssivgs}). The undisturbed population is determined by a
calculation of the population matrix $P_{ij}$ of equation
\ref{master2a} when the electrometer is switched off ($V_{b1}=800
\mu$V).} \label{table2}
\end{table}


\begin{references}
\bibitem{superdensity}  Using Fermi's golden rule we numerically calculate the
superconducting rates assuming the BCS density of states, which is
proportional to $\frac{E}{\left( E^{2}-\Delta ^{2}\right)
^{\frac{1}{2}}}$.
\end{references}
\end{document}